# SEARCH FOR EXTRA DIMENSIONS WITH ATLAS AND CMS DETECTORS AT THE LHC

S. SHMATOV

*for ATLAS and CMS Collaborations*

*Joint Institute for Nuclear Research, Dubna, Russia*

*E-mail: shmatov@cern.ch*

A brief review of the discovery potential of the ATLAS and CMS experiments to search for signals from extra dimensions in different luminosity scenarios is given.

## 1. Introduction

The Standard Model (SM) has had a tremendous success describing physical phenomena up to energies ~ 100 GeV. At present, however, there are many theoretical attempts to go beyond the SM in order to solve the known disadvantages of SM like the mass hierarchy problem, Yukawa hierarchy (explanation of mass patterns for quarks and leptons), number of generations, CP-violation problem, an arbitrariness in values of coupling constants etc. Multidimensional theories at TeV energies, as given in the brane world scenarios with large or compact extra spatial dimensions (ED), which propose a solution of the gauge hierarchy problem and some other problems, have been discussed in details.

The most experimentally-interesting feature of recent ED models is rich low-energy phenomenology that originates from the KK spectrum of various particles propagating in ED. A series of searches can be addressed to LHC experiments with different signatures.

In what follows we consider the ADD [1], RS1 [2], TeV$^{-1}$ [3], and UED [4] models and discuss various experimental signal and the discovery potential of the ATLAS and CMS experiments to observe these effects.

## 2. ADD Model

The ADD model [1] gives multiple production of KK modes of graviton that leads to violations from SM predictions in the TeV scale region. The characteristic feature of ADD is the existence of light KK-gravitons which could be directly produced at colliders (real graviton production) in $q\bar{q} \to gG$, $gq \to qG$ $gg \to gG$ processes or observed through virtual KK-graviton exchange. In these cases experimental signals for the ADD scenario might be found in dijet, dilepton, and diphoton mass spectra as well as missing energy distributions.

**Real graviton production.** Real gravitons carry away a fraction of the total energy produced in a hard collision, in other words induce energy leakage from the interaction point accompanied by a mono-jet or photons. The model with up to four extra dimensions could be probed in the channel with mono-jets at LHC (Figure 1). For 100 fb$^{-1}$, the maximum reach in a fundamental Planck mass, $M_D$, is between 9.1 TeV ($\delta = 2$) and 6.0 TeV ($\delta = 4$).

In the $G_{ADD}+\gamma$ channels a 5 σ discovery can be made with less than 1 fb$^{-1}$ of data for scenarios with $M_D$ in the range of 1-1.5 TeV,

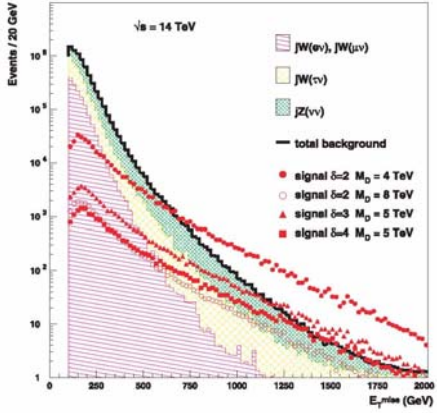

Fig. 1. Missing energy distribution (*dots*), shown here for various choices of the number of ED ($\delta$) and of the mass scale ($M_D$) and for SM backgrounds (*histograms*) [5]

and less than 10 fb$^{-1}$ for values of $M_D$ in the range of 2-2.5 TeV, largely independent of the number of extra dimensions. These estimates [6] are conservative taking into account only the events for which the graviton mass is smaller than $M_D$ and should be considered as lower bound. The discovery reach for ADD extra dimensions via this channel with 60 fb$^{-1}$ is about 3-3.5 TeV.

**Virtual graviton production.** Virtual gravitons interfere with the SM diagrams, as an example, for Drell-Yan processes as well as for gamma pair production, which results in significant modification of these spectra. This kind of signatures is clear, very sensitive to new physics and could signal the existence of extra dimensions.

The reachable values of the fundamental Planck mass, $M_D$, with $5\sigma$ significance for various number of extra dimensions are shown in Figure 2 as a function of an integrated luminosity [7]. The uncertainties related to misalignment and trigger systematic effects, PDFs, QCD-scale errors, EW and QCD corrections were taken into account. It shows that even the first LHC run with an integrated luminosity of 1 fb$^{-1}$ allows exploration of the new ADD model scale region between 3.9 and 5.5 TeV uncovered so far by other colliders (LEP and TEVATRON). An increase of the collected luminosity up to 100 fb$^{-1}$ makes it possible to probe low-scale gravity with $M_D$ in the range 5.7–8.3 TeV. In the LHC asymptotic regime the CMS sensitivity to the fundamental

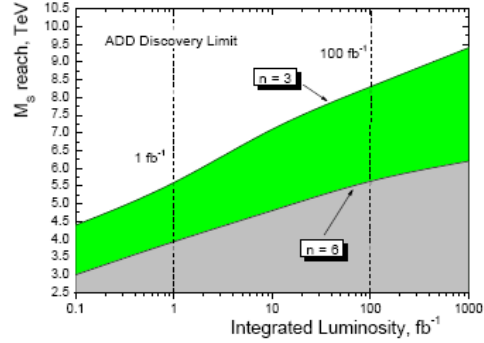

Fig. 2. $5\sigma$ limit on $M_D$ for the number of extra dimensions $n$ = 3, 4, 5, 6. Muon pair production is considered [7].

Planck scale is increased up to values of 5.9–8.8 TeV.

In the process of di-photon production ADD model could be probed up to a scale $M_D \approx$ 4.9-6.3 (6.3-7.9) TeV for low (high) luminosity, depending on the number of extra dimensions [8].

### 3. Randall-Sundrum Model (RS1)

The distinctive feature of Randall-Sundrum (RS1) approach [2] is that excited massive graviton states (RS1 graviton) are strongly coupled to ordinary particles (not suppressed below the Planckian scale like for ordinary graviton in usual description of gravity) and can significantly contribute to the SM processes above the fundamental scale. Thus, the characteristic experimental signature for these processes is a pair of high-$p_T$ leptons, gammas or jets coming from the same vertex.

**Dilepton states.** The Figures 3 shows the LHC discovery potential for Randall-Sundrum graviton in the both dilepton channels (dielectron and dimuon) for the different value of integrated luminosities [9, 10]. These results summarized a full simulation and reconstruction analyses which were carried out taking into account systematic uncertainties. About 0.8-2.3 TeV mass region is available for exploration during the LHC startup run with 1 fb$^{-1}$ in the di-muon channel for $c = 0.01$-$0.1$. For higher statistics, 10 fb$^{-1}$, a RS1 graviton can be observed with a 5σ limit up to the mass values of 1.2–3.1 TeV and 1.35-3.3 TeV in the dimuon and dielectron channel, respectively.

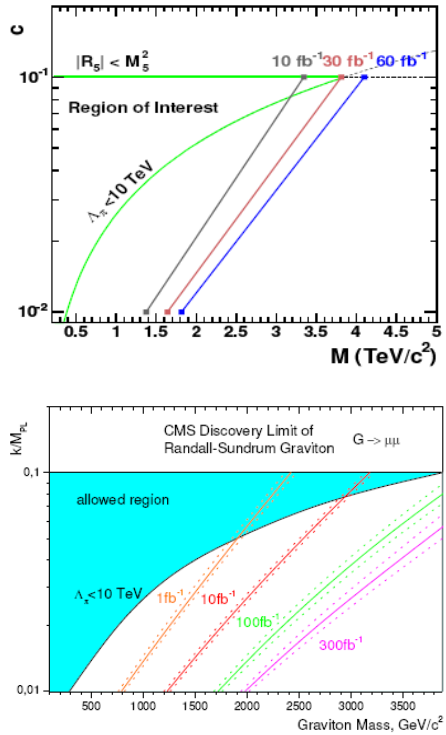

Fig. 3. Theoretical and LHC constraints on the RS1 scenario parameter in the $k=M_{Pl}$ and $m_1 = m_G$ plane for di\electrons (top) [9] and dimuons (bottom) [10].

**Diphoton states.** The overall 5σ limit for diphoton channel with a 30 fb$^{-1}$ integrated luminosity is 1.61 for low coupling $c = 0.01$, and 3.95 TeV for high coupling = 0.1 [11].

Once a new resonance has been discovered, its properties have to be determined, in order to establish its identity. A characteristic feature of the graviton is its spin-2 nature. Other particles usually considered which give a similar signature are spin-1 $Z^0$ extra gauge bosons or Kaluza-Klein excitations of a Z boson. The spin of the observed resonance manifests itself in the angular distributions of its decay products, $\cos\theta^*$, where $\theta^*$ is the angle between the incident quark or gluon and the outgoing lepton (photon or jet) in the dimuon center-of-mass frame. For an integrated luminosity of $Ldt = 100$ fb$^{-1}$, the spin-1 and spin-2 hypotheses can be discriminated at 2σ level for graviton decay into a muon pair with mass up to 1.1 TeV for $c = 0.01$ and 2.5 TeV for $c = 0.1$ [10].

## 4. TeV$^{-1}$ Extra Dimension Model

The gravitons are not the only particles possibly sensitive to extra dimensions. For example, gauge bosons could propagate in TeV$^{-1}$-sized extra dimensions [3]. The phenomenological consequence of this scenario is the appearance of the KK tower of states of gauge boson fields. The LHC will provide sufficiently high energy in the centre of mass that allows the CMS experiment to search for the direct production of new heavy resonances. The main discovery channel comes from the observation of the decay of a heavy particle into an electron and muon pair ($Z_{KK}$ decays), which presents a clear resonance signature over a well controlled Drell-Yan background or lepton an missing $E_T$ carried out by neutrinos ($W_K$ decays).

**$Z_{KK}$ decays.** The 5σ discovery limit is given as a function of the mass in Figure 4 for Kaluza-Klein excitations of Z decaying into an electron-positron pair [9]. For $Z_{KK}$ boson production, a 5σ discovery limit is achieved for compactification scales up to $M_C$ = 4.97 TeV for an integrated luminosity of 10 fb$^{-1}$, $M_C$ = 5.53 TeV for 30 fb$^{-1}$ and M = 5.88 TeV for 60 fb$^{-1}$.

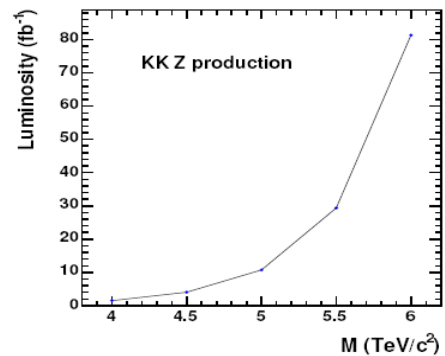

Fig. 4. Five σ discovery limit for $Z_{KK}$ boson production [9].

**$W_{KK}$ decays.** To find the of the KK excitations of the $W$ boson the leptonic signatures, $W_{KK} \to l^{\pm}\nu$, can be used. With an integrated luminosity of 100 fb$^{-1}$ a peak in the lepton-neutrino invariant transverse mass (Figure 5) can be detected if the compactification scale is below 6.0 TeV [12]. Even in absence of a peak, a detailed study of the transverse mass shape will allow to observe a deviation from the Standard Model due to the interference of the KK excitations with the SM bosons. From a study based on a maximum likelihood estimation of the compactification mass, a signal for $M_C$ < 11.7 TeV will be able to exclude at 95% CL, for one lepton flavour and with an integrated luminosity of 100 fb$^{-1}$, neglecting systematic effects.

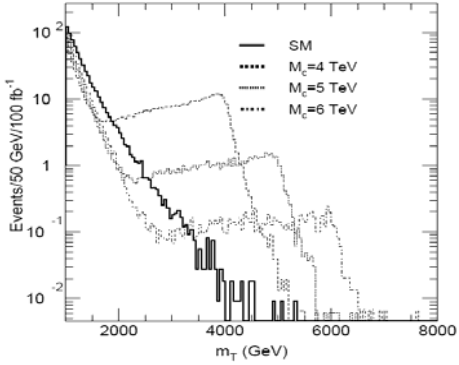

Fig. 5. Invariant transverse mass distribution of e$^{\pm}\nu$, for different values of the compactification scale $M_c$ [12].

**Gluon excitation decays.** KK excitations of gluon fields are manifested itself via decays: $q\bar{q} \to g^* \to b\bar{b}$, $q\bar{q} \to g^* \to t\bar{t}$. The analysis of expected signals and backgrounds [13] shows that the decays into $b^*$ quarks are difficult to detect, but the decays into $t$-quarks, on the contrary, might yield a significant signal for the g* mass below 3.3 TeV.

## 5. Universal Extra Dimensions

The Universal Extra Dimensions model (UED) [4] is a multidimensional field theory in which all the Standard Model (SM) fields, fermions as well as bosons, propagate in the bulk, so that each SM particle has the infinite tower of Kaluza-Klein (KK) partners.

In particular, the minimal UED (mUED) assumes that the fields can propagate in a single extra dimension. In this case the 1$^{st}$ level KK states must be pair produced and the lightest massive KK particle (LKP) is the KK photon and it is stable.

The experimental signatures for KK modes production at hadron colliders are the missing energy carried away by the LKPs in addition to soft leptons and/or jets radiated in the cascade decay process. In Figure 6, the required integrated luminosity for a 5σ significance (reference value and value including systematics) is also shown. A discovery of mUED physics with $R^{-1}$ = 300 and 500 GeV will be possible with a luminosity below 1 fb$^{-1}$, which corresponds roughly to three months LHC running at $L = 10^{33}$ cm$^{-2}$ s$^{-1}$.

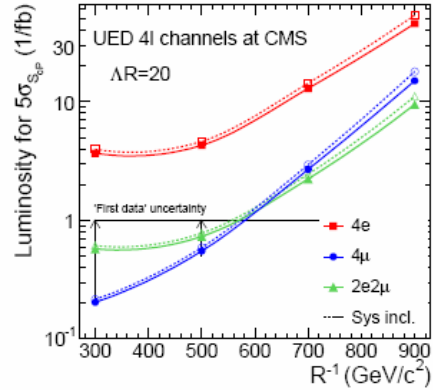

Fig. 6. Discovery potential of UED signals in 4$l$ channel. The doted lines show the influence of experimental uncertainties and the background cross-section uncertainty [14].

The highest significance levels are achieved for the channels with muons because the trigger and reconstruction efficiencies are larger than for electrons, specially for low transverse momentum. For $R^{-1} = 300$ GeV the lepton spectrum is quite soft and in consequence the $4\mu$ channel is the most effective. Above $R^{-1}$ 600 GeV the $2e2\mu$ channel becomes the most sensitive because its cross section is two times larger than for the $4\mu$ channel and it is not significantly affected by the trigger inefficiency for low momentum electrons.

If the SM brane is endowed in the ED has a finite thickness then the 1st KK states decay into a graviton and SM particle. The signature expected for such models is therefore a dijet plus a large amount of a missing $E_T$. For example, if $M_{KK} = 1.3$ TeV these UED can be probed with only 6 pb$^{-1}$ luminosity, which is indeed a clear signal. With an integrated luminosity of 100 fb$^{-1}$, a $5\sigma$ discovery is allowed if the mass of the first excited KK partons is < 2685 GeV (i.e. it is sensitive to a compactification radius of the extra dimension of ~ 2.7 TeV$^{-1}$) [15].

## Acknowledgments

We thank our colleagues from the ATLAS and CMS Collaborations for their advice and fruitful discussions, in particular B. Cousins, I. Golutvin, A. Lanyov, M.-C. Lemaire, M. Savina, M. Spiropulu, P. Traczyk, S. Valuev.